\documentclass[reprint,
superscriptaddress,
 amsmath,amssymb,
 aps,
floatfix,
]{revtex4-1}
\usepackage{graphicx}
\usepackage{dcolumn}
\usepackage{bm}
\usepackage[colorlinks]{hyperref}
\usepackage{xcolor}
\usepackage{enumerate}

\usepackage{multirow}

\newcommand{\abs}[1]{\lvert #1 \rvert}

\begin{document}

\title{Machine learning the thermodynamic arrow of time}
\author{Alireza Seif}

\affiliation{Department of Physics, University of Maryland, College Park, MD 20742} 
\affiliation{Joint Quantum Institute, NIST/University of Maryland, College Park, MD 20742}

\author{Mohammad Hafezi}

\affiliation{Department of Physics, University of Maryland, College Park, MD 20742} 
\affiliation{Joint Quantum Institute, NIST/University of Maryland, College Park, MD 20742}
\affiliation{Department of Electrical and Computer Engineering, University of Maryland, College Park, Maryland 20742, USA}

\author{Christopher Jarzynski}
\affiliation{Department of Physics, University of Maryland, College Park, MD 20742} 
\affiliation{Department of Chemistry and Biochemistry, and 
Institute for Physical Science and Technology, University of Maryland, College Park, Maryland 20742, USA}

\date{\today}

\begin{abstract}
The mechanism by which thermodynamics sets the direction of time's arrow has long fascinated scientists. Here, we show that a machine learning algorithm can learn to discern the direction of time's arrow when provided with a system's microscopic trajectory as input. The performance of our algorithm matches fundamental bounds predicted by nonequilibrium statistical mechanics.  Examination of the algorithm's decision-making process reveals that it discovers the underlying thermodynamic mechanism and the relevant physical observables. Our results indicate that machine learning techniques can be used to study systems out of equilibrium, and ultimately to uncover physical principles.

\end{abstract}

\maketitle


\section{\label{sec:intro}Introduction}
While the microscopic dynamics of physical systems are time reversible, the macroscopic world clearly does not share this symmetry. 
If we are shown a video of a macroscopic process, it is often easy to guess whether the movie is played in the correct or in time-reversed order, as in the latter case the observed sequence of events is utterly implausible.
In 1927, Sir Arthur Eddington coined the phrase ``time's arrow" to express this asymmetry in the flow of events, arguing that it traces back to the second law of thermodynamics \cite{eddington28}.

In recent decades there has been increased interest in the out-of-equilibrium physics of microscopic systems, leading to a deepened understanding of non-equilibrium fluctuations and their relation to the second law~\cite{jarzynski2011,seifert2012}.
In particular, it has become appreciated that fluctuations lead to an effective ``blurring'' of time's arrow at the nanoscale, and that our ability to discern its direction can be quantified in a system-independent manner~\cite{feng2008length,jarzynski2011,hofmann2017heat}.


Simultaneously, there have been significant advances in the ability of machine learning (ML) and artificial intelligence (AI) algorithms to tackle practical problems and to automate useful tasks. These include image and video classification~\cite{krizhevsky2012imagenet,simonyan2014very,szegedy2015going,KarpathyCVPR14,pickup2014seeing,wei2018learning}, medical diagnosis~\cite{esteva2017dermatologist}, playing games~\cite{silver2016mastering}, driving cars~\cite{buehler2009darpa}, and most recently analyzing scientific problems such as protein folding \cite{alphafoldblog}. ML methods have also emerged as exciting tools to study problems in statistical and condensed matter physics, such as classifying phases of matter, detecting order parameters, and generating configurations of a system from observed data~\cite{torlai16,carrasquilla2017machine,van2017learning,deng2016exact,wetzel17,wetzel2017unsupervised,khatami17,khatami18,liu2018discriminative,schindler2017probing,arsenault2014machine,beach2018machine,vannieuwenburg18,ponte17}. These studies extend and further motivate the use of computer algorithms to learn physics from big data~\cite{schmidt2009distilling,rudy2017data}. 

In the present work we ask whether a machine can learn to accurately guess the direction of time's arrow from microscopic data, and if so, whether it does so by effectively discovering the underlying thermodynamics, identifying relevant quantities such as work and entropy production. We approach this problem within the framework of nonequilibrium statistical mechanics, numerically generating microscopic trajectories of irreversible physical processes.  
In many of the examples we consider, the system is small and the direction of time's arrow is blurred, in the sense that both a given trajectory and its time-reversed image represent plausible sequences of events.
In these cases the algorithm in principle cannot be perfectly accurate, and it becomes interesting to ask whether it is able to assess its own likelihood to guess the direction of time's arrow correctly.

We find that the machine not only correctly classifies the direction of time's arrow but also approximates the likelihood in the uncertain cases. Moreover, the machine can generate representative trajectories for forward and backward time directions correctly, i.e., it learns what a forward/backward trajectory should look like. We also design a neural network that can detect the underlying process and classify the direction of time's arrow at the same time. 
Finally, we look inside the machine's decision-making process and find that it correctly identifies dissipated work as the key quantity for optimally guessing the direction of time's arrow.

We first introduce the relevant physical laws governing microscopic, non-equilibrium fluctuations. We then briefly review the ML techniques that we will use. Finally, we apply our methods to various model physical examples and we study the ability of ML techniques to learn and quantify the direction of time's arrow.
\begin{figure*}
    \centering
    \includegraphics[width=\linewidth]{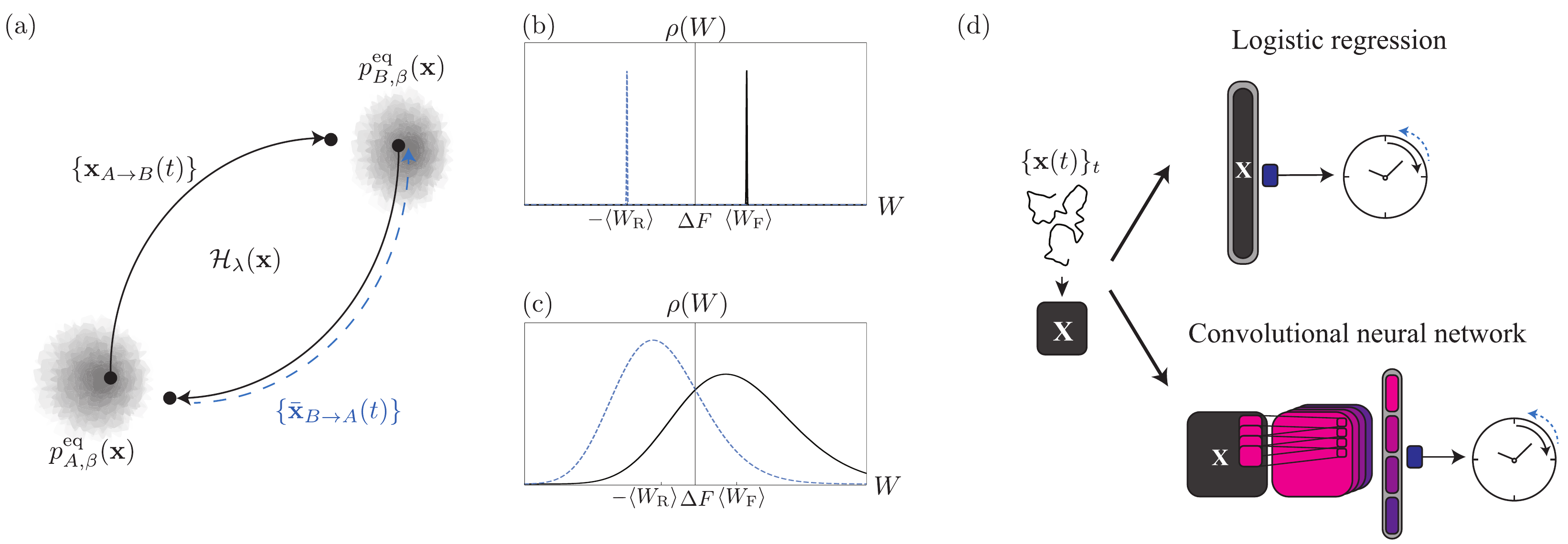}
    \caption{Non-equilibrium physics,  time's arrow, and machine learning. (a) The system evolves under a Hamiltonian that depends on an externally controlled parameter $\lambda$. The solid black trajectories depict the system's evolution during the forward and reverse process.  The dashed blue trajectory $\{\bar{\mathbf{x}}_{B\to A}(t)\}$ is the time-reversal of the system's evolution during the reverse process. (b) The distribution of work values corresponding to the forward $W_{\rm{F}}$ (solid black)  and the backward $-W_{\rm{R}}$ (dashed blue) trajectories. For macroscopic irreversible phenomena, fluctuations are negligible, $W_{\rm{F}}>\Delta F>-W_{\rm{R}}$, and the distinction between the forward and backward trajectories are clear. (c) A schematic distribution of work values in forward (solid black) and backward (dashed blue) trajectories in a microscopic system. Fluctuations are more pronounced in this case, and the distinction between the two distributions is less clear. (d) A trajectory is represented by a matrix $\mathbf{X}$ (dark grey squares). This matrix is the input to a neural network which detects the direction of the time's arrow. The top shows logistic regression network, where the input is flattened and reshaped into a vector (vertical grey rectangle), and the output (dark blue) is calculated by applying a non-linear function to a linear combination of the input coordinates. The bottom shows a convolutional neural network, where at first filters (small pink squares) are convolved with the input, making feature maps (large shades of pink squares) that encode abstract information about the local structure of the data. Then these feature maps are reshaped (vertical grey rectangle) and processed through a fully-connected layer (dark blue). The output of the network is used to decide the direction of time's arrow.}
    \label{fig:Fig1}
\end{figure*}

\section{Thermodynamics and the arrow of time\label{sec:statmech}}

When small systems undergo thermodynamic processes, fluctuations are non-negligible and the second law is expressed in terms of averages.
Thus the Clausius inequality relating the work $W$ performed on a system to the net change in its free energy, $\Delta F$, takes the form 
\begin{equation}
    \langle W \rangle \geq \Delta F,
\end{equation}
where the angular bracket denotes an average over many repetitions of the process.
Moreover, these non-equilibrium fluctuations satisfy strong constraints that allow us to rewrite such inequalities in terms of stronger equalities \cite{jarzynski1997neq,jarzynski1997eq,crooks1998,crooks1999entropy,hummer2001}, and to quantify the direction of time's arrow as a problem in statistical inference~\cite{crooks1998,shirts03,maragakis2008bayesian,feng2008length,jarzynski2011}. To frame this problem, let us first specify the class of processes we will study, and introduce notation.

Consider a system in contact with a thermal reservoir at temperature $\beta^{-1}$. The system's Hamiltonian $\mathcal{H}_\lambda(\mathbf{x})$ depends on both the system's microstate  $\mathbf{x}$, and on an externally controllable parameter $\lambda$. An external agent performs work by manipulating this parameter. Now imagine that the system begins in equilibrium with the reservoir, and the agent then varies the parameter according to a schedule $\lambda_{\rm{F}}(t)$ from $\lambda_F(0)=A$ to $\lambda_F(\tau)=B$. We refer to this as the {\it forward process}. The trajectory describing the system's evolution can be pictured as a movie, and is denoted by $\{\mathbf{x}_{A\to B}(t)\}$, where the time interval $0\leq t\leq \tau$ is implied. We also imagine the {\it reverse process}, in which the system starts in an equilibrium state at $\lambda=B$, and the agent varies the parameter from $B$ to $A$ according to $\lambda_{\rm{R}}(t) = \lambda_{\rm{F}}(\tau -t)$. The trajectory (movie) for this process is denoted by $\{\mathbf{x}_{B\to A}(t)\}$. Finally, consider the time reversal of this trajectory, $\bar{\mathbf{x}}_{B\to A}(t)= \mathbf{x}^*_{B\to A}(\tau-t)$, where the `$*$' implies negation of momentum coordinates. This time-reversed trajectory corresponds to a movie of the reverse process, played backward in time; the same trajectory may have been achieved during a realization of the forward process, see Fig.~\ref{fig:Fig1}(a).

Throughout this paper, we will use the term {\it forward trajectory} to refer to a trajectory generated during the forward process, i.e.\ $\{\mathbf{x}_{A\to B}(t)\}$, and we will use the term {\it backward trajectory} to denote a trajectory generated during the reverse process, but run backward in time, i.e.\ $\bar{\mathbf{x}}_{B\to A}(t)$, depicted by the dashed blue line in Fig.~\ref{fig:Fig1}(a).

Guessing the direction of time's arrow can be cast as a game in which a player is shown either a forward or a backward trajectory -- thus in either case the player ``sees'' the parameter being varied from $A$ to $B$.
The player must then guess which process, forward or reverse, was actually used to generate the trajectory~\cite{jarzynski06}. The player's score, or accuracy, is the ratio of correct predictions to the total number of samples.

In order to optimize the likelihood of guessing correctly, it suffices for the player to know the sign of the quantity $W - \Delta F$, where $W$ is the work performed on the system and $\Delta F = F_B - F_A$ is the free energy difference between its initial and final states, as depicted in the movie. Specifically, let $P({\rm{F}}|\{\mathbf{x}(t)\})$ denote the likelihood that a given trajectory, $\{\mathbf{x}(t)\}$, is obtained by performing the forward process, and let $P({\rm{R}}|\{\mathbf{x}(t)\})$ denote the likelihood that the trajectory is the time reversal of a realization of the reverse process. Note that $P({\rm{F}}|\{\mathbf{x}(t)\})+P({\rm{R}}|\{\mathbf{x}(t)\})=1$.  In addition, assume that the game is unbiased, e.g.\ the choice of performing the forward or reverse process in the first place was decided by flipping a fair coin.
Then the likelihood that the trajectory was generated during the forward process is given by~\cite{shirts03,maragakis2008bayesian,jarzynski2011}
\begin{equation}
\label{eq:likelihood}
    P({\rm{F}}|\{\mathbf{x}(t)\}) = \frac{1}{1+e^{-\beta(W-\Delta F)}},
\end{equation}
which is greater than (less than) 50\% when $W - \Delta F$ is positive (negative).
Here, the work performed by the external agent is  
\begin{equation}
\label{eq:defW}
W = \int_0^\tau dt \dot{\lambda} \frac{\partial \mathcal{H}_\lambda(\mathbf{x})}{\partial \lambda},
\end{equation}
and the change in free energy is given by 
\begin{equation}
    \Delta F = -\frac{1}{\beta}\log\left(\frac{Z_{B,\beta}}{Z_{A,\beta}}\right),
\end{equation}
where 
\begin{equation}
Z_{\lambda,\beta} = \int d\mathbf{x} \exp[-\beta\mathcal{H}_\lambda(\mathbf{x})]
\end{equation}
is the partition function. 

In macroscopic systems, the values of work performed on the system corresponding to forward trajectories, $W_{\rm{F}}$, and for backward trajectories, $-W_{\rm{R}}$, are sharply peaked around their mean values, Fig.~\ref{fig:Fig1}(b), and the sign of $W-\Delta F$ is a reliable indicator of the direction of time's arrow. (Here, $W_R$ is the work performed during a given realization of the reverse process, therefore for the corresponding backward trajectory the work value is $-W_R$.)
However, for microscopic systems these distributions can overlap significantly, as in Fig.~\ref{fig:Fig1}(c).
Eq.~\ref{eq:likelihood} shows that the player optimizes the chance of success simply by guessing ``forward'' whenever $W > \Delta F$, and ``reverse'' otherwise, without accounting for any further details of the trajectory.
Note that if $\abs{W - \Delta F} \gg k_BT$ then determining the arrow of time is easy, but when $\abs{W - \Delta F} \lesssim k_BT$ the problem becomes more difficult -- in effect, time's arrow is blurred.

\section{Neural networks} 
\label{sec:neuralnets}

We wish to train a computer program to infer the direction of time's arrow from a movie of the system's trajectory.
To do so, we first simulate a number of trajectories from the forward and the reverse processes, and we ``time-reverse'' the latter so that each trajectory is chronologically ordered with $\lambda$ varying from $A$ to $B$. We attach a label $y=0$ (reverse) or $y=1$ (forward) indicating which process was used to generate that trajectory.  We then provide the machine with this collection of labelled trajectories, which will serve as the training data.
{\it A priori}, any one of the trajectories could have been generated from either the forward or the reverse process, and the training stage now consists of using a neural network (NN) to construct a model of the function $P({\rm{F}}|{\mathbf{x}(t)})$, which gives the likelihood the trajectory was generated by the forward process.
Although this function is known analytically, Eq.~\ref{eq:likelihood}, the machine is not provided with this information.
We now sketch how the training is accomplished.

Since each (numerically generated) trajectory consists of a discretized time series of microstates, we represent the trajectory as a matrix $\mathbf{X}$ whose rows correspond to different times, and whose columns correspond to phase space coordinates.
The training stage amounts to designing a function that maps any such matrix $\mathbf{X}$ onto a real number $p$ between 0 and 1, whose value is the machine's best estimate of the likelihood that the trajectory was generated by the forward process.

In this work, we consider two types of classifiers: (i) logistic regression (LR), and (ii) convolutional neural network (CNN). The input to LR is a vectorized trajectory $\mathbf{a} = {\rm{vec}}(\mathbf{X})$, and the output is $p = g(\mathbf{\Omega}^\intercal \mathbf{a} + b)$, where $\mathbf{\Omega}$ is a vector of weights, $b$ is the bias, and $g(z) = 1/(1+\exp(-z))$ is the logistic sigmoid function, see the top panel of Fig.~\ref{fig:Fig1}(d). The CNN can compute more complicated functions than the LR \cite{Goodfellow-et-al-2016}. The input to our CNN is a trajectory matrix $\mathbf{X}$, and the output is again a value $p$. The CNN has convolutional layers that extract useful information by taking advantage of the temporal and spatial structure of the data, see the bottom panel of Fig.~\ref{fig:Fig1}(d). For details of the CNN architecture see Appendix~\ref{app:nndetails}.

To train the network, we determine optimal values of parameters (such as the weights and biases in LR) by minimizing the cross-entropy\cite{Goodfellow-et-al-2016} 
\begin{equation}
\label{eq:cost}
    C=-\frac{1}{N_{\rm samp}}\sum_m \left[ y_m \log(p_m) + (1-y_m)\log(1-p_m) \right] \, .
\end{equation}
Here the sum is carried over the $N_{\rm samp}$ training samples, $y_m \in \{0,1\}$ is the label attached to the $m$'th trajectory, indicating which process was actually used to generate the trajectory, and $p_m$ is the output of the network for that trajectory.

Throughout this work, we always split a given data-set into three parts. We use 60\%, 20\% and 20\% of the data for training, validation, and testing the model, respectively. The validation set is used to tune the architecture and hyperparameters of the model, while the test data is used for unbiased evaluation of the  the final model's accuracy. We use Adam optimizer with parameters suggested in the original paper for the training \cite{kingma2014adam}. We assess the performance of the network by testing it over a balanced set of trajectories, i.e. half forward and half backward. If  $p_m\geq0.5$ then the algorithm guesses that the trajectory was generated from the forward process, otherwise it guesses the reverse process.
As a figure of merit, we consider the accuracy, i.e. the ratio of correct guesses to total number of samples. 
The best score that an algorithm can achieve, in the limit of a very large test set, is obtained if the output of the network agrees with the theoretical likelihood  \eqref{eq:likelihood}, in other words if the algorithm ``learns'' a result from nonequilibrium statistical physics.

For additional considerations in training NNs see Appendix~\ref{app:nndetails} 

\section{Case studies\label{sec:casestudies}}
We apply the neural network machinery to detect the direction of the time's arrow and assess the NN's accuracy. We also look at the output of the network and compare it with the theoretical optimal result of Eq.~\eqref{eq:likelihood}. Interestingly, the networks not only learn to guess the direction of the time's arrow but also learn to closely reproduce the likelihood function (see Appendix~\ref{app:altact} for a discussion of the sensitivity of the results to the choice of the activation functions).

We first consider a single Brownian particle in a moving potential. This problem is simple and has an analytical solution. 

We then move on to the more complicated problem of a spin chain with nearest-neighbour coupling in a magnetic field. We consider two scenarios involving the spin chain. First, the coupling is assumed to be constant, and the magnetic field is varied in time. Next, the magnetic field is constant the coupling is changed through time. We refer to the former as the $\textsf{B}$ protocol, and the latter as the $\textsf{J}$ protocol.  

For details of the numerical calculations used to generate the trajectories see Appendix~\ref{app:gendata}. For details about the NNs and the number of samples used see Appendix~\ref{app:nndetails}. Table~\ref{tab:perf} summarizes the accuracy of the algorithms studied in this and the following sections.

\subsection{Brownian particle in a moving potential \label{sec:bpart}}
\begin{figure}[t]
    \centering
    \includegraphics[width=\columnwidth]{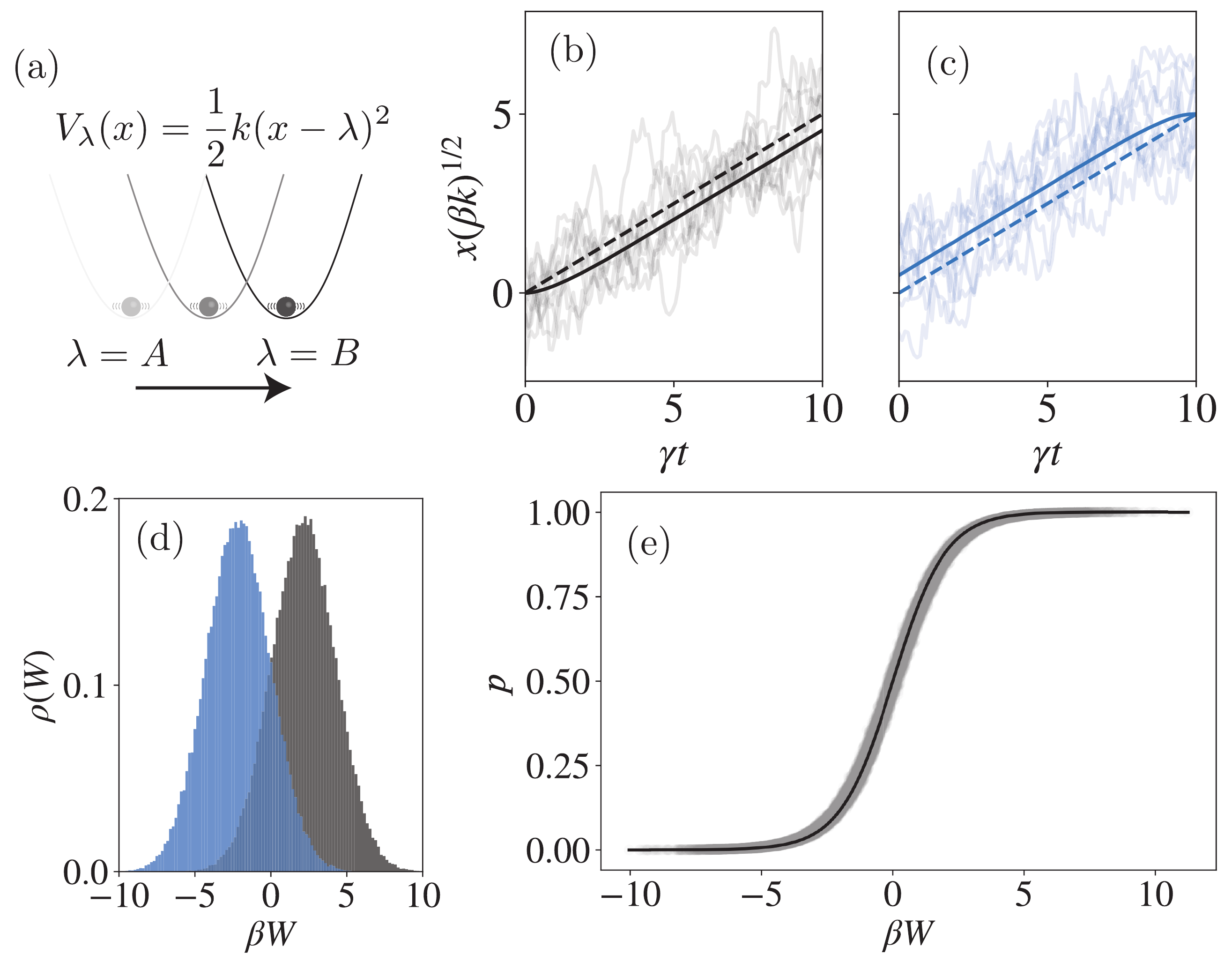}
    \caption{ Brownian particle in a moving potential. (a) An overdamped Brownian particle at temperature $\beta^{-1}$ is in a harmonic potential $V_\lambda(x)$, with stiffness $k$. The position of the potential's center, $\lambda$, is externally controlled and is moved from $A$ to $B$ in the forward process. (b) Sample trajectories (grey) and the average trajectory (black) in the forward protocol. Note that the average trajectory lags behind the center of the potential (dashed line). (c) Sample backward trajectories (light blue) and their average (dark blue) in the reverse process. The average trajectory leads the potential's center (dashed line). (d) Work distribution for the forward (black) and the backward (blue) trajectories. They are both normally distributed and are symmetric around 0. (e) The likelihood of the forward process for a set of test trajectories. The output of the neural network, $p$, over the test set (grey circles) resembles the theoretical $P({\rm{F}}|\mathbf{X})$ (solid black line).}
    \label{fig:Fig2}
\end{figure}

Consider an overdamped Brownian particle at temperature $\beta^{-1}$ in a harmonic potential (see Fig.~\ref{fig:Fig2}(a)), evolving according to 
\begin{equation}
    \label{eq:brownian}
    \dot{x} = -\frac{k}{\gamma} (x-\lambda) + \xi(t), 
\end{equation}
where $k$ denotes the strength of the potential, $\lambda$ is the position of the center of the potential, and $\gamma$ is the damping rate. The noise term, $\xi(t)$ satisfies $\langle \xi(t)\xi(t')\rangle = 2 (\beta \gamma)^{-1}\delta(t-t')$. 
In the forward protocol, the value of $\lambda$ is changed from $A$ to $B$ at a fixed rate $\dot{\lambda}=u$. Hence the reverse protocol changes $\lambda$ from $B$ to $A$ with $\dot{\lambda}=-u$.  
If the potential is moved rapidly, then in most of the forward trajectories the particle lags substantially behind the potential, whereas in most of the backward trajectories the particle leads the potential. In these cases, the direction of time's arrow is clear. However, if the potential is moved slowly, then the particle stays near the center of the potential, the processes approach the reversible limit, and it becomes difficult to determine the direction of time's arrow.  The theoretical likelihood   \eqref{eq:likelihood} is determined by the work $W$ performed and the free energy change, $\Delta F$. Note that in this protocol $\Delta F = 0$. For each trajectory, we calculate $W$ by integrating  
\begin{equation}
\label{eq:workparticle}
\dot{W} = -k u (x-ut),
\end{equation}
We generate samples of the forward and backward trajectories by numerically integrating the stochastic differential equation Eq.~\eqref{eq:brownian} (see Fig.~\ref{fig:Fig2}(b) and (c)). We then train a classifier to predict the label for a given trajectory, as described earlier. In Fig.~\ref{fig:Fig2}(e) we compare the accuracy and the output of a LR classifier (grey circles) with the theoretical likelihood (solid curve) obtained from  Eqs.~\eqref{eq:workparticle} and \eqref{eq:likelihood}, see Table~\ref{tab:perf} and Fig.~\ref{fig:Fig2}(e). 

The seemingly remarkable agreement with the theory can be understood by examining $\dot{W}$ \eqref{eq:workparticle}. Namely, the work $W$ calculated by numerically integrating $\dot{W}$ for a given trajectory is linearly related to the sum of the components of that vector. Therefore, LR is well-equipped to calculate this quantity and reproduce the likelihood function. See Appendix~\ref{app:optimalBP} for detailed analysis of the optimal network. 

\subsection{Spin chain - time-dependent field  \label{sec:spinB}}
\begin{figure*}[t]
    \centering
    \includegraphics[width=.8\linewidth]{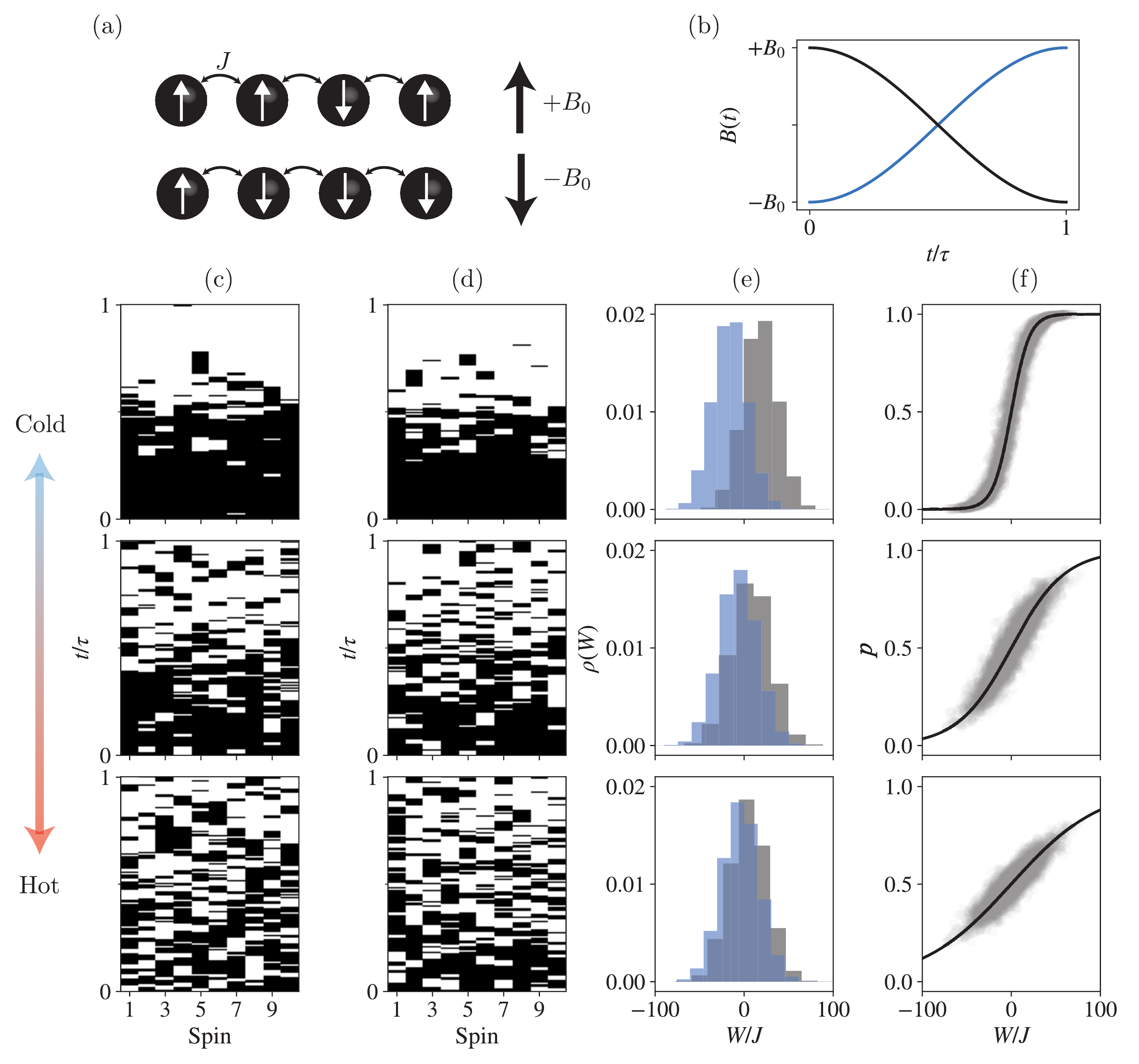}
    \caption{Spin chain in a time-dependent magnetic field. (a) A chain of ten spins with periodic boundary condition is placed in a magnetic field. The strength of coupling between nearest neighbors is $J$. The forward process starts with spins in equilibrium at temperature $\beta^{-1}$ with $B=+B_0>0$ and ends at a non-equilibrium state with $B=-B_0$. (b) The forward (black) and the reverse (blue) protocols $B(t)$. (c) Sample forward and (d) sample backward trajectories, where the black and white pixels denote spins pointing up and down, respectively.  (e) The distribution of work for the forward (black) and backward (blue) trajectories, (f) the theoretical likelihood function (solid black line) and the output of the neural network over the test set (grey circles)  for various temperatures. In this example, a single network is trained simultaneously with trajectory data with different $\beta$ values. The temperatures corresponding to different rows in panels (c), (d), (e), and (f) correspond to $\beta^{-1}/J = 10, 30, 50$ in descending order. As the temperature increases, the distinction between the forward and backward trajectories is blurred.  In these simulations $B_0/J = 20$. }
    \label{fig:Fig3}
\end{figure*}

Now let us consider a more complicated, many-particle system and a non-linear work protocol. Specifically, we consider a spin chain in a time-dependent magnetic field $B(t)$ and in contact with a thermal reservoir at temperature $\beta^{-1}$, see Fig.~\ref{fig:Fig3}(a), described by a Hamiltonian
\begin{equation}
\label{eq:spinhamB}
    H = \sum_i J \sigma_i \sigma_{i+1} - B(t) \sum_i \sigma_i,
\end{equation}
where $\sigma_i \in\{-1,+1\}$ is the spin variable at site $i$, and $J$ is the nearest-neighbour coupling strength.  The dynamics of this system are modeled as a Markov process (See Appendix.~\ref{app:gendata}). The Hamiltonian aligns the spin in preferred energy configurations, while thermal fluctuations cause the spins to flip randomly according to a rate related to $\beta$. We refer to this example as the $\mathsf{B}$ protocol.

In the forward process, $B(t)$ changes from a positive value $B_0$ at $t=0$ to a negative value $-B_0$ at $t=\tau$, as shown in Fig.~\ref{fig:Fig3}(b). In the limit where $B_0/J \gg \beta^{-1}/J \gg 1$, the spins start mostly aligned with the magnetic field. As the field magnitude is lowered, thermal fluctuations become dominant and the spins flip randomly. Eventually, $B(t)$ becomes large and negative, and aligns the spin in the other direction (see the top row of Figs.~\ref{fig:Fig3} (c) and (d)). In this limit it is easy to detect the direction of the time's arrow, as the work distributions have a modest overlap, see Fig.~\ref{fig:Fig3}(e) top row. As the temperature is increased, thermal fluctuations increase the overlap in work distributions, blurring the direction of time's arrow, see the middle and the bottom rows of Figs.~\ref{fig:Fig3}(c), (d), and (e). 

To train the classifier, we generate samples of forward and backward trajectories for three different temperatures using the Metropolis algorithm. The trajectories are matrices with $\pm1$ entries, whose rows and columns correspond to time steps and spin positions, respectively. We are interested in training a single LR classifier that is capable of detecting the direction of time's arrow for different temperatures. Therefore, the information about $\beta$ is provided through normalizing the elements of the trajectory data with their corresponding temperatures. This matrix is then reshaped as a vector to serve as the input to an LR classifier. We observe that the success of LR in learning both the correct labels and in approximating the likelihood function persists, see Table~\ref{tab:perf} and Fig.~\ref{fig:Fig3}(f).  The reason, again, lies in the functional form of $W$, which can be evaluated by numerically integrating
\begin{equation}
    \label{eq:workspinB}
    \dot{W} = -\dot{B}(t)\sum_i \sigma_i.
\end{equation}
It can be seen that $W$ is proportional to the weighted sum of the elements of the input vector. Note that, in this protocol $\Delta F =0$. Consequently, LR is a perfect model of the likelihood function for all the temperatures, see Appendix~\ref{app:optimalBP}.

\subsection{Spin chain - time-dependent coupling \label{sec:spinJ}}

\begin{figure*}[t]
    \centering
    \includegraphics[width=.8\linewidth]{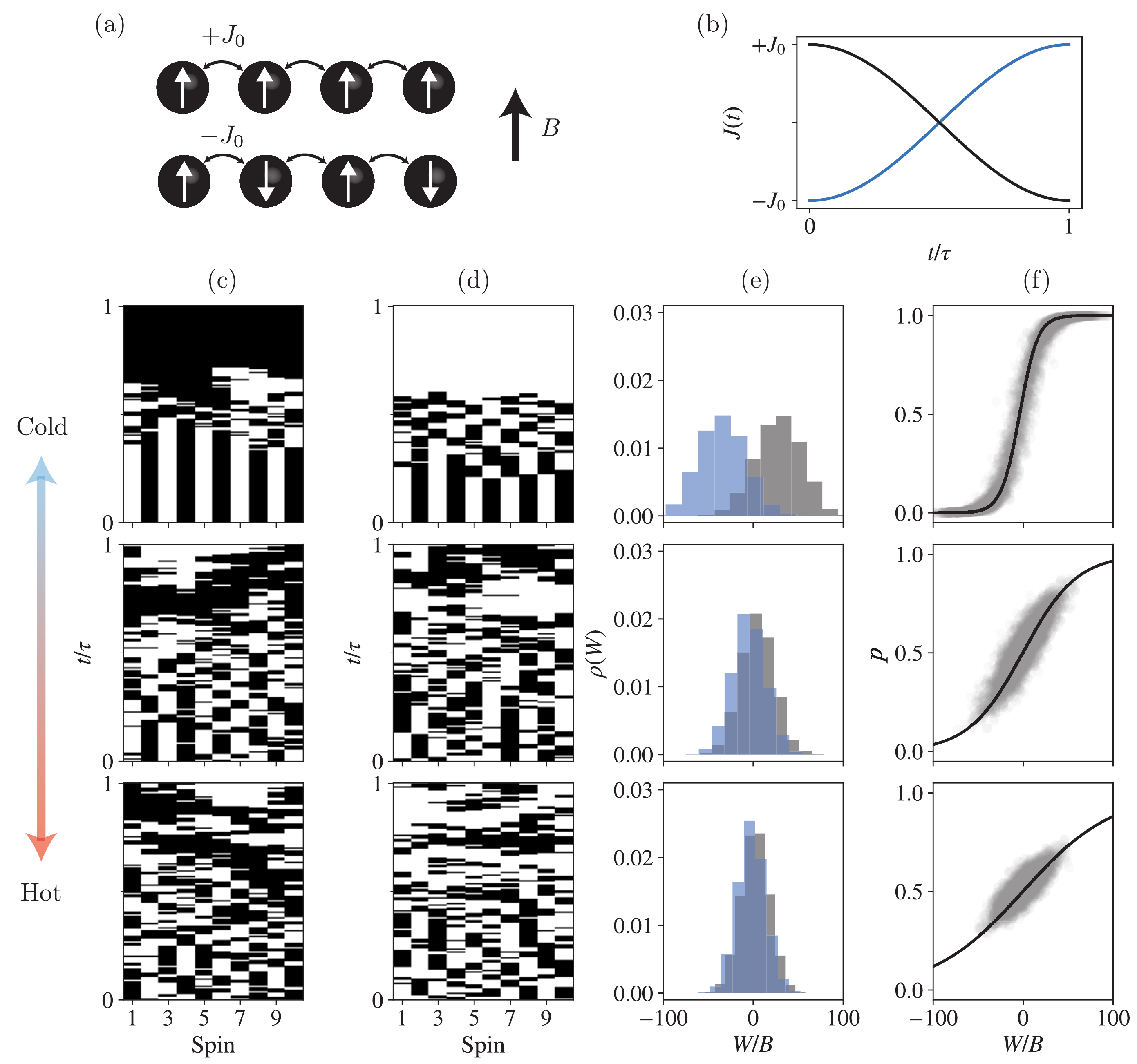}
    \caption{Spin chain with a time-dependent coupling. (a) A chain of ten spins with periodic boundary condition is placed in a constant magnetic field $B$. The time-dependent coupling between nearest neighbors is $J(t)$. The forward process starts with spins in equilibrium at temperature $\beta^1$ with $J(0)=+J_0>0$ and ends at a non-equilibrium state with $J(\tau)=-J_0$. (b) The forward (black) and the reverse (blue) protocols $J(t)$. (c) Sample forward and (d) sample backward trajectories, where the black and white pixels denote spins pointing up and down, respectively.  (e) The distribution of work values for the forward (black) and backward (blue) trajectories, (f) the theoretical likelihood function (solid black line) and the output of the neural network over the test set (grey circles) for various temperatures. In this example, a single network is trained simultaneously with trajectory data with different $\beta$ values. The temperatures corresponding to different rows in panels (c) - (f) correspond to $\beta^{-1}/B = 10, 30, 50$ in descending order. As the temperature increases, the distinction between the forward and backward trajectories is blurred In these simulations $J_0/B = 20$.}
    \label{fig:Fig4}
\end{figure*}

In this example, we consider a more complicated version of the spin chain problem, with a ferromagnetic-antiferromagnetic transition. Here, we keep $B$ constant and positive and allow for the time-dependent couplings $J(t)$, see Fig.~\ref{fig:Fig4}(a). We refer to this example as the $\mathsf{J}$ protocol.  The Hamiltonian is given by
\begin{equation}
\label{eq:spinhamJ}
    H = \sum_i J(t) \sigma_i \sigma_{i+1} - B \sum_i \sigma_i.
\end{equation}

The protocol $J(t)$ is shown in Fig.~\ref{fig:Fig4}(b). In the forward case  $J(t)$ is varied from $J_0>0$ at $t=0$ to a $-J_0$ at $t=\tau$. Note that at low temperatures, where $J_0/B\gg\beta^{-1}/B\gg1$, the spins start in a state with anti-ferromagnetic ordering. As $J(t)$ grows weaker during the protocol, thermal fluctuations dominate. By the end of the protocol, $J(t)=-J_0$ and the system settles in a state with ferromagnetic ordering. In this case, the forward and reverse work distributions are distinguishable, and so is the arrow of time, see the top rows of Fig.~\ref{fig:Fig4}(c) - (e). As the temperature increases, so does the overlap between the distributions (the bottom two rows of Fig.~\ref{fig:Fig4}(c) - (e)). 

In this case the work is given by the time integral of
\begin{equation}
    \label{eq:workspinJ}
    \dot{W} = \dot{J}(t)\sum_i \sigma_i \sigma_{i+1}.
\end{equation}
We see that $W$ is no longer linearly related to the input, and the LR classifier is incapable of calculating it. Therefore, we use a CNN with periodic boundary condition that can capture more complicated functions. With this CNN, we are able to recover the optimal accuracy again. Note that in this process $\Delta F\neq0$, which adds another layer of complexity to the problem. The convolution layer in a CNN has filters that can capture the two-body nearest-neighbor correlations required to calculate the work, without introducing too many parameters. In fact, for a single temperature, we are able to analytically derive the parameters of a CNN that exactly calculate the likelihood function. The performance and the output of the network are shown in Table~\ref{tab:perf} and Fig.~\ref{fig:Fig4}(f).   For more details on the optimal network construction and the performance of sub-optimal strategies see Appendix~\ref{app:optimalSpinJ}. 

\begin{table}[t]
    \centering
    \begin{tabular}{|c|c|c|}
    \hline
        Example & Accuracy (theory) &  Accuracy (NN) \\
        \hline \hline
        Brownian particle & $84\%$ & $84\%$  \\
        \hline
         Spin chain $\mathsf{B}$ & $(81\%,63\%,58\%)$ & $(80\%,61\%,57\%)$  \\
        \hline
        Coarse-grained $\mathsf{B}$ &  $(81\%,63\%,58\%)$ &  $(81\%,62\%,57\%)$\\
        \hline
        Spin chain $\mathsf{J}$ (LR) & \multirow{3}{*}[0.5em]{$(89\%,60\%,56\%)$} & $(67\%,50\%,50\%)$\\
        \cline{1-1}\cline{3-3}
         Spin chain $\mathsf{J}$ (CNN) && $(89\%,59\%,54\%)$\\
        \hline 
        Coarse-grained $\mathsf{J}$ & 89\% & 88\%\\
    \hline
    \end{tabular}
    \caption{Comparison of the accuracy of the neural networks with the theoretical optima. The numbers in a tuple denote the accuracy of the corresponding NN at different temperatures. In these cases, the networks are simultaneously trained at different temperatures. For the \textsf{B} protocol they correspond to $\beta^{-1}/J = 10, 30, 50$, respectively. Similarly, they correspond to $\beta^{-1}/B = 10, 30, 50$ for the \textsf{J} protocol. }
    \label{tab:perf}

\end{table}

\section{Interpretation and extensions \label{sec:insidenn}}
In this section, we use three approaches to investigate trained networks and to develop insight into what they have learned. 
\begin{figure}[h]
    \centering
    \includegraphics[width=1\columnwidth]{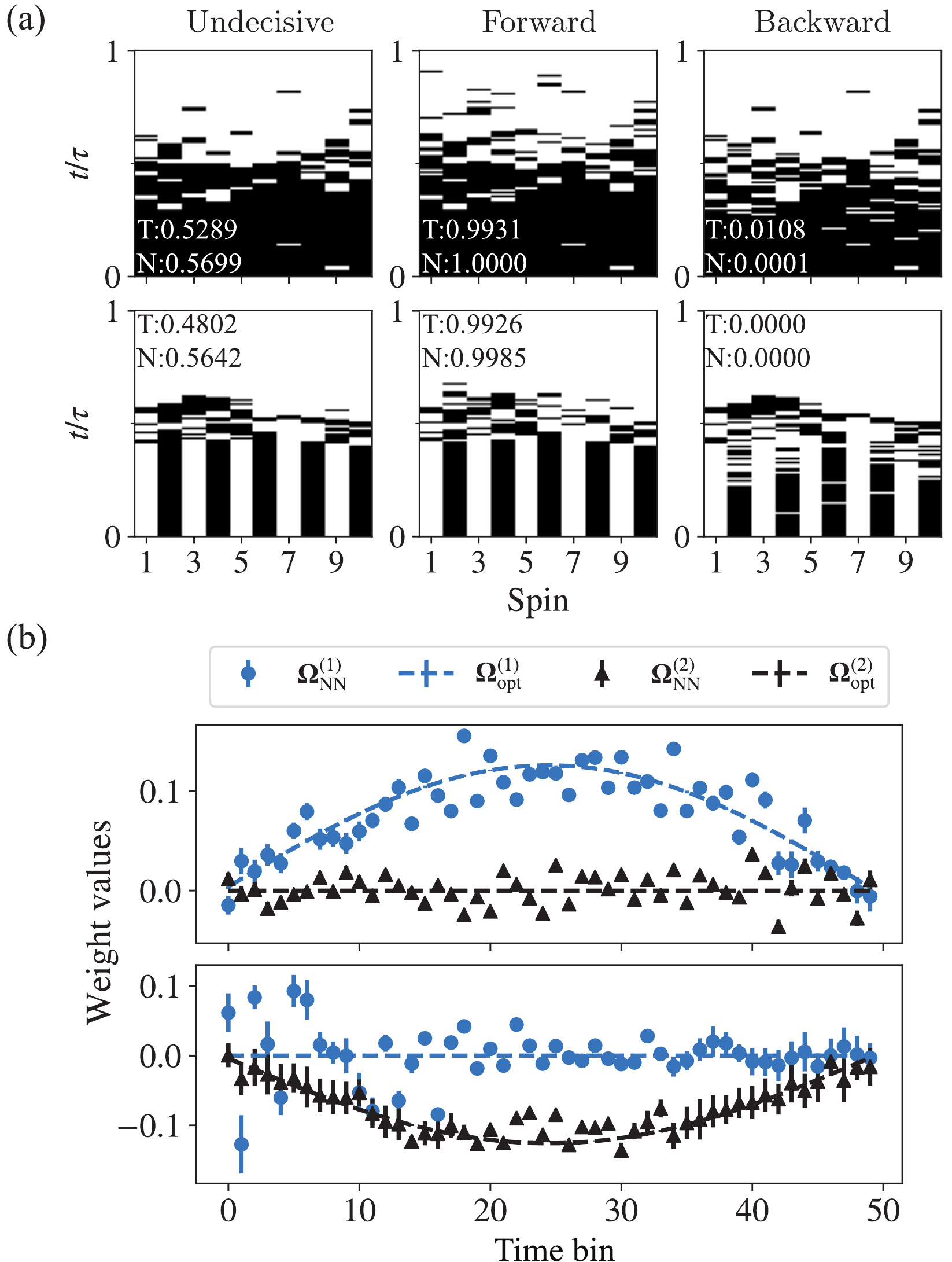}
    \caption{Interpreting the neural network's inner mechanism. 
    (a) Starting with a random trajectory (leftmost column), we ask the network to `dream' of its idea of the forward (middle column) and backward (rightmost column) trajectories. The top row corresponds to the \textsf{B} protocol, and the bottom row corresponds to the \textsf{J} protocol. The black and white pixels denote spins pointing up and down, respectively. The numbers in the inset indicate the forward likelihood  $P({\rm{F}}|\mathbf{X})$, obtained from the theory (T) using Eq.~\eqref{eq:likelihood} and from the neural network's output (N). 
    (b) The weights of the network associated with the magnetization  $\mathbf{\Omega}^{(1)}$ and the nearest-neighbour correlations  $\mathbf{\Omega}^{(2)}$ for the \textsf{B} protocol (top row) and the \textsf{J} protocol (bottom row). The error bars are standard deviation over 10 trained networks with random weight initialization. The network bases its decision on the magnetization in the former, and on the nearest-neighbor correlations in the latter case. If the values of  $\mathbf{\Omega}^{(\ell)}_{\rm{NN}}$s of the trained networks (markers) match the optimal weights $\mathbf{\Omega}^{(\ell)}_{\rm{opt}}$ (dashed line), the output of the network agrees with the exact likelihood \eqref{eq:likelihood}.}
    \label{fig:Fig5}
\end{figure}

First, we use inceptionism techniques \cite{inception15,neupert17} to learn the network's ideal representative of forward and backward trajectories. 
 Specifically, we use gradient descent on a random input such that the trained networks in Secs.~\ref{sec:spinB} and \ref{sec:spinJ} output 1 or 0 corresponding to forward and backward trajectories, respectively. This is in contrast with the previous section where we optimized for the weights and biases of the network. Among the simulated trajectories in the test set, we choose one with $p\approx0.5$ \cite{neupert17} -- this is a trajectory for which the classifier has difficulty assigning the direction of time's arrow. We project the configurations to discrete values after each step of the gradient descent, and demand that there be at most 1 spin-flip per time step, to ensure that the network `dreams' of physically realizable trajectories. We find that the networks' ideas of the forward and backward trajectories show strong agreement with the true physical picture, see Fig.~\ref{fig:Fig5}(a). 

Secondly, to assign a physical interpretation to the networks' decision-making process, we project the trajectories onto a two-dimensional reduced phase space corresponding to the collective coordinates $\{\tilde{x}^{(1)}(t)\} = \{\sum_i \sigma_i(t)\}$ and $\{\tilde{x}^{(2)}(t)\} = \{\sum_i \sigma_i(t)\sigma_{i+1}(t)\}$ (taking period boundary conditions), representing  magnetization and nearest-neighbour correlations, respectively.
We also replace the value of $\tilde{x}^{(\ell)}(t)$ within each time window of ten time steps, by the sum of the values within that window.
By thus coarse-graining in both phase space and time, we reduce the noise due to finite size effects and variations over samples.
Next, we use these coarse-grained trajectories to train LR classifiers for both protocols in Secs.~\ref{sec:spinB} and \ref{sec:spinJ} (See Table~\ref{tab:perf} for the performance of these networks). Finally, we investigate the weights $\mathbf{\Omega}^{(\ell)}$ that the networks assign to the magnetization ($\ell=1$) and the nearest neighbor correlations ($\ell=2$).  Fig.~\ref{fig:Fig5}(b) reveals that for the \textsf{B} protocol (top row), the network mostly cares about the magnetization, whereas when the \textsf{J} protocol is performed (bottom row), the network bases its decision on the nearest-neighbor correlations. Moreover, the learned values of $\mathbf{\Omega}^{(\ell)}$ agree with our analytical results that reproduces the correct likelihood value (see Appendices~\ref{app:optimalNN} and~\ref{app:optimalSpinCG} for details). These observations suggest that the network learns that the time derivative of the Hamiltonian, and by extension the work \eqref{eq:defW}, is an important feature in guessing the direction of time's arrow. We note that when the process is highly irreversible, the distributions of the forward and reverse work are well-separated. In this case, the network easily determines the arrow of time, but does not learn about the importance of work and bases its decision on other visible differences in the trajectories, see Appendix~\ref{app:irrev}.

\begin{figure}[h]
    \centering
    \includegraphics[width=1\columnwidth]{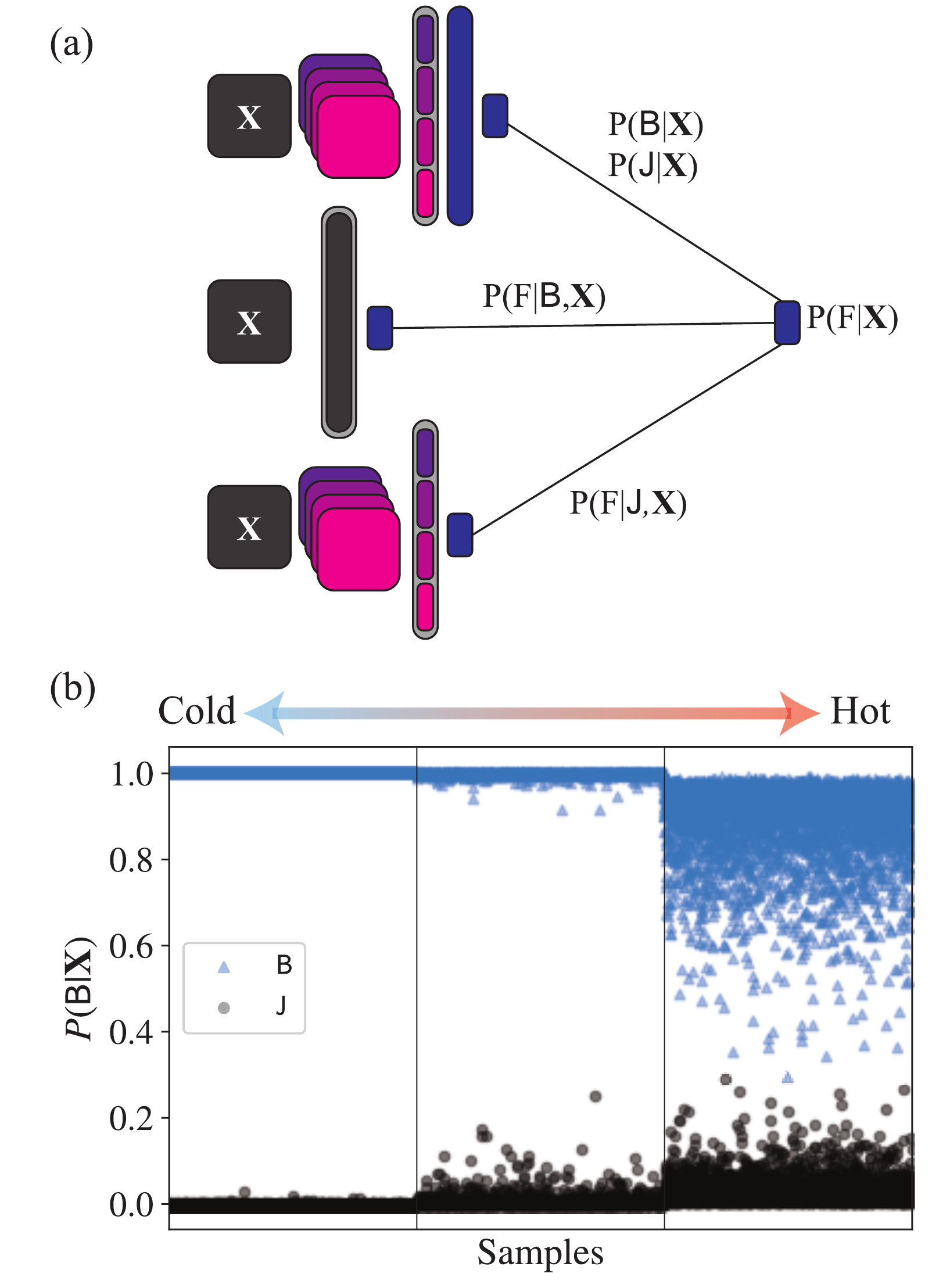}
    \caption{Mixture of experts. (a) The MoE network models the forward likelihood $P({\rm{F}}|\mathbf{X})$. It consists of a gating CNN that predicts the protocol $P(\mathsf{B}(\mathsf{J})|\mathbf{X})$, and two networks that predict the forward likelihood of a trajectory given the protocol $P({\rm{F}}|\mathsf{B}(\mathsf{J}),\mathbf{X})$. (b) The output of the gating network, which models $P(\mathsf{B}|\mathbf{X})$, is shown for different sample trajectories of the \textsf{B} (blue triangles) and \textsf{J} (black circles) protocols. The horizontal axis shows different samples in three temperature regions separated by vertical line, corresponding to $\beta^{-1}$ values in Figs.~\ref{fig:Fig3} and ~\ref{fig:Fig4}. It is harder to predict the protocol at higher temperatures.}
    \label{fig:Fig6}
\end{figure}

Lastly, we ask whether a single algorithm can learn to accurately guess the direction of time's arrow for trajectories generated using multiple protocols, when the identity of the protocol is not specified.
One approach is to take a large neural network and train it on trajectories from both $\mathsf{B}$ and $\mathsf{J}$ protocols. We have found that this approach works to a certain degree, but does not reach the accuracy of the individual networks in Secs.~\ref{sec:spinB} and \ref{sec:spinJ}. However, by using our knowledge about the structure of the problem we can design an algorithm that attains the optimal performance. Specifically, we use a  mixture of experts (MoE), with an output that is the weighted sum of expert networks \cite{nowlan1990mixexp}. When the protocol is not specified, the net forward likelihood is
\begin{equation}
\label{eq:moe}
P({\rm{F}}|\mathbf{X}) = P({\rm{F}}|\mathbf{X},\mathsf{B})P(\mathsf{B}|\mathbf{X})+P({\rm{F}}|\mathbf{X},\mathsf{J})P(\mathsf{J}|\mathbf{X}).    
\end{equation}
The quantities $P(F|\mathbf{X},\mathsf{B})$ and $P(F|\mathbf{X},\mathsf{J})$ are modeled using neural networks similar to those considered in Secs.~\ref{sec:spinB} and \ref{sec:spinJ}, respectively. These networks are referred to as experts.  Additionally, we use a CNN  to model $P(\mathsf{B}|\mathbf{X}) = 1 -P(\mathsf{J}|\mathbf{X}) $. This CNN, which is called the gating network, learns the protocol from trajectories. Therefore, we obtain a larger three-headed network by combining the output of the three neural networks as in Eq.~\eqref{eq:moe}, as illustrated in Fig.~\ref{fig:Fig6}(a). For the training, we use the pre-trained expert networks for the $\mathsf{B}$ and $\mathsf{J}$ protocols, and optimize the cost function \eqref{eq:cost} over sample trajectories from both protocols. We observe that the performance of this network is similar to that of the individual networks, as the gating network learns to accurately identify the protocol of input trajectories (see Fig.~\ref{fig:Fig6}(b)). Note that the predictions of the gating network are more accurate at lower temperatures. This makes sense as the distribution of the initial state in the two protocols are distinguishable in low temperatures, but become less so as the temperature is increased.  

\section{Conclusion and outlook}
Starting with a simple, solvable harmonic oscillator model, then proceeding to more complicated spin systems, we have shown that machine learning algorithms can be trained to discern the direction of time's arrow in irreversible thermodynamic processes.
We have found that neural networks not only learn to guess the direction of time's arrow but also to accurately evaluate the likelihood that the guess is correct, when the direction of the arrow is not entirely clear.
Moreover, we have used various techniques to interpret what the network learns.
In particular, by examining the optimized parameter values that emerge from the training, we have been able to identify which physical quantities the network uses to guess the direction of time's arrow.
In this sense, our study represents a step toward AI driven discovery of physical concepts.

Machine learning techniques have been applied extensively to the study of equilibrium statistical physics \cite{torlai16,carrasquilla2017machine,van2017learning,wetzel2017unsupervised,wetzel17,khatami17,khatami18,liu2018discriminative,schindler2017probing,beach2018machine,ponte17}.
Our results extend this computational toolkit to out-of-equilibrium phenomena.
While we have focused on the arrow of time, we expect that other important issues and questions in non-equilibrium physics can usefully be studied with these tools.
We anticipate that the techniques considered in this work can be extended to estimate free energy differences, as well as to identify physical quantities that distinguish different regimes of dynamics in out-of-equilibrium quantum phenomena.
In addition, using unsupervised learning techniques such as generative modeling may be especially useful in studying non-equilibrium phenomena \cite{torlai16,morningstar2017deep}. Unlike the equilibrium case where the state of the system is given by the Boltzmann distribution, the general form of the non-equilibrium steady-state is not known. Generative models are an ideal candidate to model and learn these distributions.

Moreover, machine learning researchers have shown that ML techniques can be used to detect the playback direction of real-world videos ~\cite{pickup2014seeing,wei2018learning}. These studies are concerned with videos of macroscopic objects that are in principle irreversible, and the arrow of time has a clear direction. In such scenarios, there are many indicators that can reveal the true playback direction, and therefore it is hard to quantify the optimal performance. However, in the physical examples the optimal attainable accuracy of the classifier is dictated by the laws of physics. Therefore, problems with large number of phase-space coordinates and with complicated dynamics, such as the \textsf{J} protocol for 2D Ising model, can serve as a standardized benchmark for video classification algorithms.   

\section*{Acknowledgements}
AS thanks Evert van Nieuwenburg, Grant Rostkoff, and Ali Izadi Rad for helpful discussions. AS and MH gratefully acknowledge support from ARO-MURI and Physics Frontier Center by National Science Foundation at the JQI and CJ from the National Science Foundation under grant DMR-1506969.

\bibliographystyle{apsrev4-1}
\bibliography{arrowoftime}

\pagebreak
\appendix

\section{Neural networks \label{app:nndetails}}
\subsection{Convolutions}
A convolution layer convolves the input with a number of filters, and then applies a non-linear function to the output of the filters. Each convolution operation with a kernel $\mathbf{\Omega}$ and bias $b$, maps an input matrix $\mathbf{X}$,  to another matrix $\mathbf{Z} = \mathbf{\Omega}*\mathbf{X}$ given by \cite{Goodfellow-et-al-2016}
\begin{equation}
    Z_{j,k}= \sum_{m,n} X_{j\times s+m,k\times s+n} \Omega_{m,n} + b 
\end{equation}
where $s$ specifies the number of steps the filter moves in each direction. It is called the stride of the convolution and is a hyperparameter that is tuned using the cross-validation data. The output of the convolution layer is obtained by applying a non-linear function $g$ element-wise to $\mathbf{Z}$. The convolution layers can be repeated many times, and combined with pooling layers where the dimension of the output is reduced through a procedure such as averaging. At the end, the output of the convolution layer is flattened to form a vector and that vector is fed into a series of fully connected layers to produce the network's output \cite{Goodfellow-et-al-2016}.

The CNN's that we consider has four $2\times2$ filters, with the stride of 1, and with periodic boundary condition. We choose the rectifier, i.e., $g(z) = \max(0,z)$, for the activation of these filters. The output of all the filters is then combined to form a single vector. For the CNN classifying the \textsf{J} protocol (Sec.~\ref{sec:spinJ}), this vector is fed into a single neuron with sigmoid activation, whose values determine the direction of time's arrow. For the gating network (Sec.~\ref{sec:insidenn}), this vector is fed into a fully connected layer with 50 hidden neurons and the rectifier activation, followed by the output neuron with the sigmoid activation.  

\subsection{Regularization and sample size}
To reduce overfitting it is helpful to include a regularization term.  This will help to reduce the difference between the training error and the test error. We consider $L_2$ regularization $\alpha \sum_{\ell}\Omega_{\ell}^2$, that is adding the square of all the weights in the network to the cost function. The parameter $\alpha$ is a hyper-parameter of the model and is tuned using the cross-validation data. 

Additionally, in training the CNN in Sec.~\ref{sec:spinJ}, we use the dropout technique to reduce overfitting. Dropout refers to deactivating and ignoring certain neurons during the training phase. Specifically, at every training step, a random fraction of $p_{\rm{drop}}$ of neurons are deactivated \cite{srivastava2014dropout}. 

We find that the performance of our algorithms does not vary significantly with the choice of hyper-parameters. We choose $p_{\rm{drop}}=0.25$ for the dropout rate of neurons of the convolutional layer in the \textsf{J} network, and $p_{\rm{drop}}=0.5$ for the gating network. The $L_2$ regularization rates are shown in Table~\ref{tab:reg}.

\begin{table}[h]
    \centering
    \begin{tabular}{|c|c|}
    \hline
        Model & $\alpha$ \\
        \hline 
        \hline
        Brownian particle & $0.001 $\\
        \hline
        Spins $\mathsf{B}$ & $10^{-4}$\\
        \hline
        Spins $\mathsf{J}$ (LR) & $2\times10^{-5}$\\
        \hline
        Spins $\mathsf{J}$ (CNN all layers) & $10^{-4}$\\
        \hline
        Spins coarse-grained (all cases) & $2\times10^{-5}$\\
        \hline
        Gating network (conv. and the hidden layer) & $10^{-5}$\\
        \hline
        Gating network (output) & $2\times10^{-5}$\\
        \hline
        Alternative activation functions (all layers) & $10^{-5}$\\
    \hline

    \end{tabular}

    \caption{The value of $L_2$ regularization parameter for the NNs in this work.}
        \label{tab:reg}

\end{table}
Another important quantity in training the neural networks is the sample size. We use a total of 20000 samples for the Brownian particle. For the spin chain examples (\textsf{B} and \textsf{J} protocols), we use 20000 samples for each temperature. The samples are then split into three sets and are used to train, validate, and test the models.

\section{Generating the data \label{app:gendata}}
To generate trajectories we closely follow Ref.~\cite{crooks1998}. We consider a discrete set of time steps $t\in\{0,1,\dots,\tau\}$. The value of the control parameter and the state of the system at each time step is denoted by $\lambda_t$ and $\mathbf{x}_t$, respectively. Note that in dealing with discrete time steps, rather than using $u(t)$, we use the notation $u_t$ for the value of variable $u$ at the time step $t$. In the forward process, the initial state of the system is drawn from equilibrium with $\lambda = \lambda_0$. The time evolution can be broken into two substeps: 
\begin{enumerate}[(i)]
    \item With the state of the system fixed, the control parameter is changed $\lambda_t \to \lambda_{t+1}$
    \item At fixed $\lambda_{t+1}$, the state of the system evolves $\mathbf{x}_t \to \mathbf{x}_{t+1}$
\end{enumerate}
Here, the second substep is either generated by a stochastic differential equations (Sec.~\ref{sec:bpart}) or Metropolis algorithm (Secs.~\ref{sec:spinB} ~\ref{sec:spinJ}). 
The total work performed in this process is 
\begin{equation}
\label{eq:discW}
    W = \sum_{t=0}^{\tau-1} [\mathcal{H}_{\lambda_{t+1}}(\mathbf{x}_t) - \mathcal{H}_{\lambda_{t}}(\mathbf{x}_t) ]
\end{equation}
For producing backward trajectories, the system is initialized in an equilibrium state with $\lambda=\lambda_\tau$. The dynamics begin with a change in the system state, followed by a change in $\lambda$. In the end, the history of the system state is reversed, and the calculated work is negated to obtain backward trajectories and their corresponding work values. 
\section{Optimal networks \label{app:optimalNN}}
For some of the examples that we considered, it is possible to derive an analytical expression for the optimal weights and biases of the network. Specifically, we examine the expression that is used to calculate the work $W$ and the change in free energy $\Delta F$. Because the logistic sigmoid activation function, i.e., $g(z)=1/(1+\exp(-z))$, used for classification coincides with the form of the likelihood function \eqref{eq:likelihood} in the arrow of time problem, we are able to find the networks parameters $\{\mathbf{\Omega},\mathbf{b}\}$ that reproduce the same likelihood function.  To illustrate, consider the LR model with the output $p=1/(1+\exp(-z))$ and $z=\mathbf{\Omega}^\intercal \mathbf{a} + b$. If we find $\mathbf{\Omega}$ and $b$ such that $z = \beta(W-\Delta F)$, the output of the network $p$ correctly represent $P({\rm{F}}|\mathbf{X})$. In the following we show that when $W$ \eqref{eq:discW}, is linear in the elements of $\mathbf{x}_t$ and is subsequently linear in elements of $\mathbf{a}$, we are able to find such optimal $\mathbf{\Omega}$ and $b$.  

\subsection{Brownian particle in a moving potential\label{app:optimalBP}}
In this example, the system's state at each time step is described by a scalar $x_t$, i.e. the position of the particle. We have a total of $\tau+1$ time steps, therefore the input to the NN is a $\tau+1$ dimensional vector. The LR classifier considered here, is parameterized by a $\tau+1$ dimensional weight vector with elements $\Omega_t$ for $t=0,\dots,\tau$ and a bias $b$. Using Eq.~\eqref{eq:discW} we find
\begin{equation}
    W = \sum_{t=0}^{\tau-1} \delta\lambda(k x_t - k\lambda_t +\frac{1}{2} k \delta\lambda),
\end{equation}
where $\delta\lambda = \lambda_{t+1} - \lambda_{t}$ is independent of $t$, because the protocol is linear. Note that $\Delta F =0$ in this example. With the choice of  
\begin{align}
    \Omega_t &= \begin{cases}\beta k \delta\lambda &(t\neq\tau) \\
    0 & (t=\tau)
    \end{cases},\\
    b &=  \beta \sum_{t=0}^{\tau-1}\delta\lambda(- k\lambda_t +\frac{1}{2} k \delta\lambda),
\end{align}
we can see that $\mathbf{\Omega}^\intercal \mathbf{a} + b = \beta(W-\Delta F)$, where $(\mathbf{a})_t = x_t$.

\subsection{Spin chain - \textsf{B} protocol \label{app:optimalSpinB}}
The full trajectory of an $n$ spin system over $\tau$ time steps is represented by a $\tau \times n$ matrix $\mathbf{X}$. We denote the orientation (up or down) of the $i$th spin at time $t$ with $X_{t,i} = \pm 1$. The input to LR classifier, is a vector obtained from rearranging the trajectory matrix $\mathbf{X}$ to shape it into an $\tau n \times 1$ array. By using Eq.~\eqref{eq:discW} we find that 
\begin{equation}
\label{eq:workBtheory}
    W = -\sum_{t=0}^{\tau-1} (\delta B_t \sum_{i=1}^{n} X_{t,i}),
\end{equation}
where $\delta B_t = B_{t+1}-B_t$. Work calculated using Eq.~\eqref{eq:workBtheory} is the discrete time version of $W$ obtained from Eq.~\eqref{eq:workspinB}. In this example, $\Delta F = 0$ again, and we find that the optimal weights and bias are given by
\begin{align}
    \Omega_{t,i} &= - \beta \delta B_t \quad\quad (t\neq\tau),\\
    b &= 0,
\end{align}
where $\Omega_{\tau,i} = 0$. Note that if the input $X_{t,i}$ is scaled with $\beta$, a single LR classifier is able to reproduce the correct likelihood for different temperatures. 

\subsection{Spin chain - \textsf{J} protocol \label{app:optimalSpinJ}}
\begin{figure}[h]
    \centering
    \includegraphics[width=1\columnwidth]{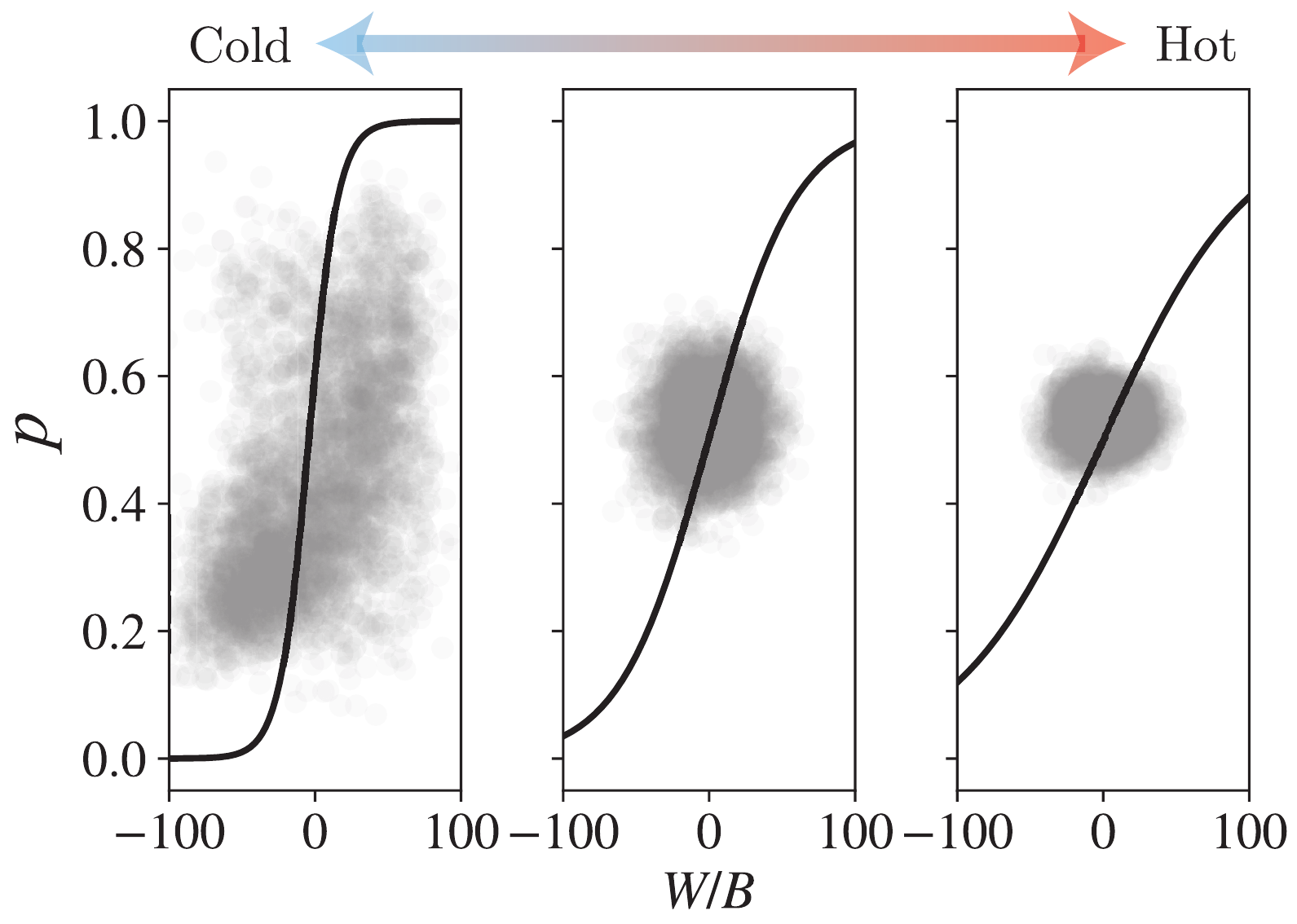}
    \caption{Performance of LR in the \textsf{J} protocol. As expected, LR does not perform well and cannot match the performance of a CNN as observed in Fig.~\ref{fig:Fig4}. The columns correspond to $\beta^{-1}/B = 10, 30,50$, respectively.}
    \label{fig:FigA1}
\end{figure}

Using the same notation in the previous section, we find that the work, $W$, is given by
\begin{equation}
\label{eq:workJtheory}
    W = \sum_{t=0}^{\tau-1} (\delta J_t \sum_{i=1}^{n} X_{t,i}X_{t,i+1}),
\end{equation}
where $\delta J_t = J_{t+1}-J_t$. Work calculated using Eq.~\eqref{eq:workJtheory} is the discrete time version of $W$ obtained from Eq.~\eqref{eq:workspinJ}.  We also use periodic boundary condition, which implies $X_{n+1} = X_1$. Moreover, there is a non-zero change in the free energy, which is given by \cite{Salinas2001}
\begin{equation}
    \Delta F = -\frac{1}{\beta}\log(\frac{\epsilon_{-}(\beta B, \beta J_0)^n+\epsilon_{+}(\beta B,\beta J_0)^n}{\epsilon_{-}(\beta B,-\beta J_0)^n+\epsilon_{+}(\beta B,-\beta J_0)^n})
\end{equation}
where 
\begin{equation}
\begin{split}
    \epsilon_{\pm}(\beta B,\beta J) &= \exp(\beta  J) \cosh (\beta  B)\\ &\pm\sqrt{\exp (2 \beta  J) \cosh ^2(\beta  B)-2 \sinh (2 \beta  J)}.
\end{split}
\end{equation}
We see that it is not possible to have a logistic regression model that calculates $W$ (see Fig.~\ref{fig:FigA1}). 

However, CNN's can, in principle, learn the relevant representation (i.e. nearest-neighbor correlations) from the input data, and learn the corresponding weights to calculate work. Specifically, to show that in principle a CNN can exactly calculate the correct likelihood from the input, we consider a CNN with four $1\times2$ filters with periodic boundary condition and the rectifier activation $g_1$, followed by the output layer with sigmoid activation $g_2$. We set the biases of the convolutional layer to zero, and choose the weights $\mathbf{\Omega}^{[1,i]}$ for filters $i=1,2,3,4$ as follows
\begin{equation}
    \begin{Bmatrix}\mathbf{\Omega}^{[1,1]} \\ \mathbf{\Omega}^{[1,2]} \\ \mathbf{\Omega}^{[1,3]} \\ \mathbf{\Omega}^{[1,4]} \end{Bmatrix}=
    \begin{Bmatrix} 
    (1 & 1)\\
    (1 & -1)\\
    (-1 & 1)\\
    (-1 & -1)
    \end{Bmatrix}.
\end{equation}
Each filter is only activated for one of the possible configuration of two neighboring spins. Specifically, given the $\tau\times n$ input $\mathbf{X}$, the output of each filter $g_1(\mathbf{\Omega}^{[1,j]} * \mathbf{X})$ is a $\tau\times n$ matrix  $\tilde{\mathbf{X}}^{(j)}$ such that 
\begin{align}
    \tilde{{X}}^{(1)}_{t,i} &= 1  \quad  \text{if }\quad (X_{t,i} X_{t,i+1}) = (1,1),  \\
    \tilde{{X}}^{(2)}_{t,i} &= 1  \quad  \text{if }\quad (X_{t,i} X_{t,i+1}) = (1,-1),  \\
    \tilde{{X}}^{(3)}_{t,i} &= 1  \quad  \text{if }\quad (X_{t,i} X_{t,i+1}) = (-1,1),  \\
    \tilde{{X}}^{(4)}_{t,i} &= 1  \quad  \text{if }\quad (X_{t,i} X_{t,i+1}) = (-1,-1),  
\end{align}
and $\tilde{{X}}^{(j)}_{t,i} = 0$ otherwise. We can now rewrite the output of the network as 
\begin{equation}
    g_2(b_2 + \sum_j (\sum_{t,i}\Omega^{[2,j]}_{t,i} \tilde{{X}}^{(j)}_{t,i}))),
\end{equation}
where $\mathbf{\Omega}^{[2,j]}$ contains the weights of the output layer corresponding to $\tilde{\mathbf{X}}^{(j)}$. 
The optimal values of these weights are given by 
\begin{align}
    \Omega^{[2,1]}_{t,i}=\Omega^{[2,4]}_{t,i} &= - \beta \delta J_t \quad\quad (t\neq\tau),\\
    \Omega^{[2,2]}_{t,i}=\Omega^{[2,3]}_{t,i}&= + \beta \delta J_t \quad\quad (t\neq\tau),\\
    b &= -\beta \Delta F,
\end{align}
where $ \Omega^{[2,j]}_{\tau,i} = 0$. 

Therefore, a CNN with four $1 \times 2$ filters is sufficient to capture the likelihood at a single temperature of $\beta^{-1}$. We find that in practice, a CNN with such an architecture is likely to get stuck at local minima, and finding the optimal parameters shown above greatly depends on the initial weights of the network. However, we observe that a CNN with four $2 \times 2$ filters can achieve a close to optimal performance more easily.

\section{Coarse-grained features \label{app:optimalSpinCG}}
\begin{figure}[h]
    \centering
    \includegraphics[width=1\columnwidth]{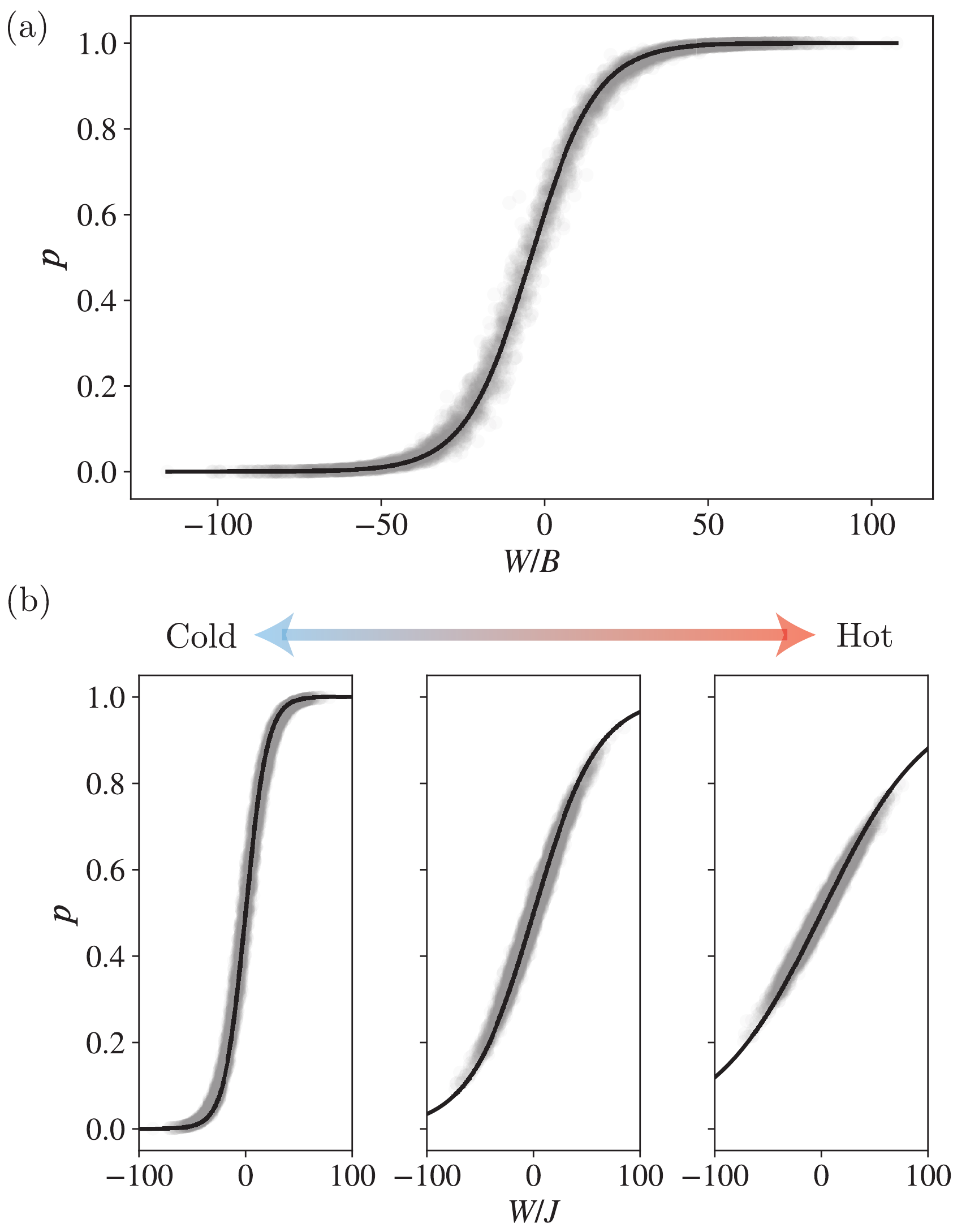}
    \caption{Prediction of LR with coarse-grained features.  The output of the network (grey circles) and the theoretical likelihood function (solid curve) agree remarkably in this case. Coarse-graining and feature engineering improves the performance. (a) LR's prediction at $\beta^{-1}/B = 10$ for the \textsf{J} protocol. (b) LR predicting the direction of time's arrow in the \textsf{B} protocol for three different temperatures corresponding to $\beta^{-1}/J = 10, 30, 50$, respectively. }
    \label{fig:FigA2}
\end{figure}
To reduce the parameters of the neural network and simplify the task of learning we pre-calculate a set of features for the network. Specifically, for the two protocols concerning the spin chain in a magnetic field, the coarse-grained features are 
\begin{align}
    \tilde{x}^{(1)}_{s} &= \sum_{t=m s}^{m (s+1)-1} \sum_{i=1}^{n} X_{t,i},\\
     \tilde{x}^{(2)}_{s} &= \sum_{t=m s}^{m (s+1)-1} \sum_{i=1}^{n} X_{t,i}X_{t,i+1},
\end{align}
where $m$ is an integer and $s$ is the scaled time. Using this feature map, LR classifier can calculate $W$ for both \textsf{B} and \textsf{J} protocols (See Fig.~\ref{fig:FigA2}). The input to the network is a $2\tau/m$ dimensional vector $\begin{bmatrix} \mathbf{\tilde{x}}^{(1)} \\ \mathbf{\tilde{x}}^{(2)}\end{bmatrix}$. We denote the weights corresponding to $\mathbf{\tilde{x}}^{(\ell)}$ by $\tau/m$ dimensional vectors $\mathbf{\Omega}^{(\ell)}$ for $\ell = 1,2$. 
In this case, we approximate the optimal weights and bias of the networks by their average over the coarse-grained time window. For the \textsf{B} protocol we find 
\begin{align}
    \Omega^{(1)}_s &= \frac{\beta }{m}\sum_{t=m s}^{m (s+1)-1} \delta B_t,\\
    \Omega^{(2)}_s &= 0,\\
    b &= 0.
\end{align}
Similarly, the weights and bias for the \textsf{J} protocol are given by 
\begin{align}
    \Omega^{(1)}_s &= 0,\\
    \Omega^{(2)}_s &= \frac{\beta }{m}\sum_{t=m s}^{m (s+1)-1} \delta J_t,\\
    b &= -\beta  \Delta F.
\end{align}
In both cases we can see that $\beta(W-\Delta F)\approx(\mathbf{\Omega}^{(1)})^\intercal\tilde{\mathbf{x}}^{(1)}+(\mathbf{\Omega}^{(2)})^\intercal\tilde{\mathbf{x}}^{(2)}$, where the approximation comes from coarse-graining. 

Note that in this case, even though LR classifier can calculate $\beta W$ if the input is scaled with $\beta$, it is not possible to train the network over different temperatures. This is because $\Delta F \neq 0$, and a simple bias cannot capture multiple values of $\beta \Delta F$. Therefore, we only consider a single temperature for the \textsf{J} protocol in studying the optimal networks with coarse-grained features.  

\section{Alternative activation functions \label{app:altact}}
\begin{figure}[h]
    \centering
    \includegraphics[width=1\columnwidth]{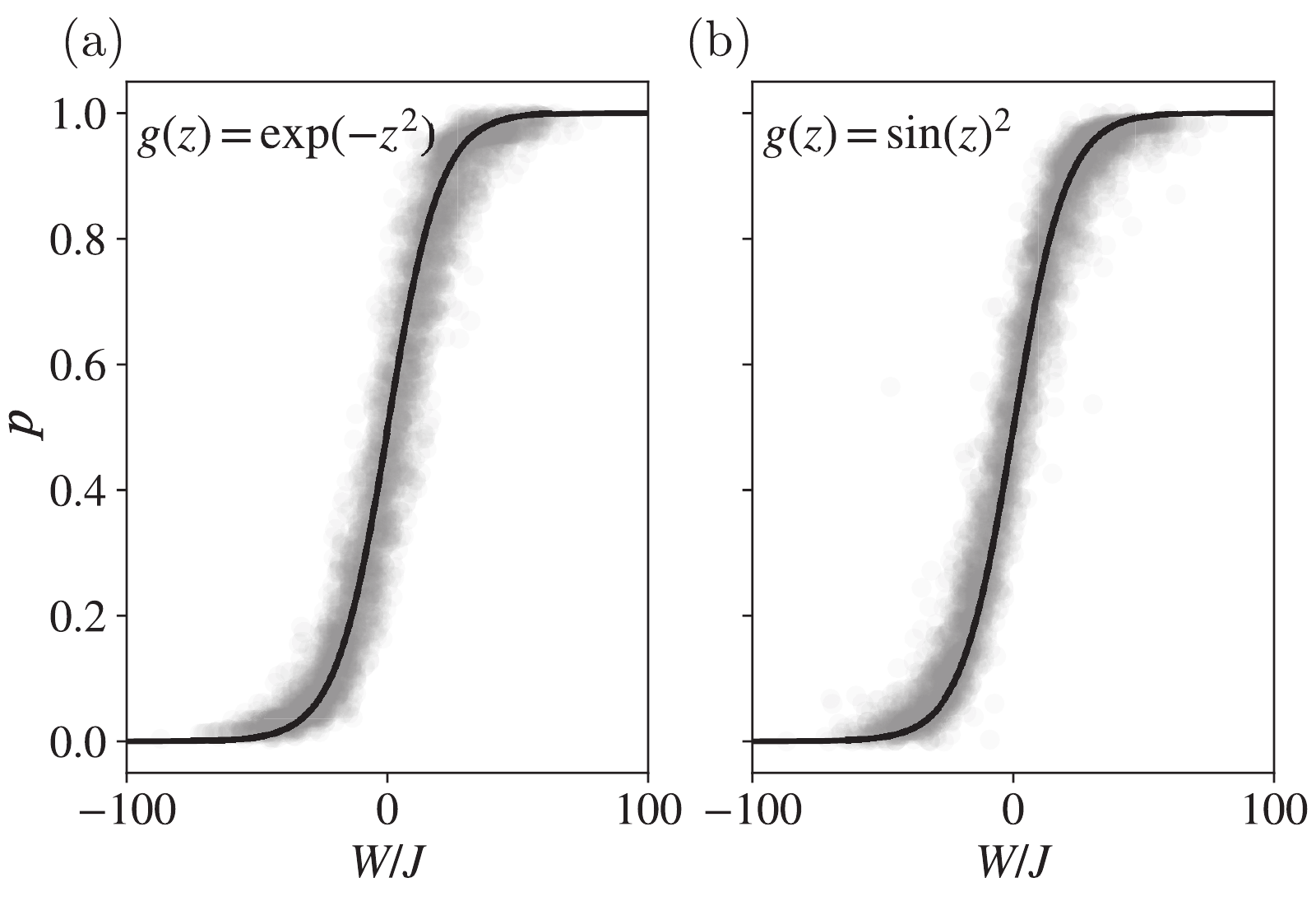}
    \caption{The output of the neural network with custom activation functions for the last layer. The network has a hidden layer with 50 neurons with $\tanh$ activation. The last layer's activation functions are (a) $g(z)=\exp(-z^2)$ and (b)$g(z)=\sin(z^2)$. Comparing the plots with the leftmost column of Fig.~\ref{fig:FigA2}, we observe that the performance deteriorates. However, the networks still capture the essence of the likelihood function.}
    \label{fig:FigA3}
\end{figure}

The logistic function that appears in the theoretically calculated likelihood in the time's arrow problem, is similar to the activation function that is commonly used for classification in machine learning. To assess the general ability of the networks in approximating the likelihoods we try different activation functions. Specifically, we choose 
$g(z) = \exp(z^2)$ and $g(z)=\sin(z)^2$ as the activation of the last layer of the neural network so that the output is always between 0 and 1, and can be interpreted as probabilities.  We also add a hidden layer to give the network the ability to calculate complex functions.
We compare the network's output with the theoretical likelihoods for the spin chain under the \textsf{B} protocol with coarse-grained feature discussed in Sec.~\ref{app:optimalSpinCG}. We only train the network at a single temperature and observe that the network can still approximate the likelihood function as shown in Fig.~\ref{fig:FigA3}. 

\section{Highly irreversible processes \label{app:irrev}}
 When the process is highly irreversible, the arrow of time has a clear direction. In such cases, we observe that while LR has 100\% accuracy, it does not learn work (obtained by inverting the sigmoid function in the output). This is because, the events that enable the network to learn work are extremely rare, and are usually absent in the training data. However, there are other evident differences that can show the direction of time's arrow. We show an example of such a process for the \textsf{B} protocol at low temperatures. We observe that the orientation of spins undergoes a sharp transition as the magnetic field changes sign. However, the time that this transition occurs is different in the forward and backward trajectories. The classifier makes a decision based on the spin configuration at this particular time.
\begin{figure}[h]
    \centering
    \includegraphics[width=1\columnwidth]{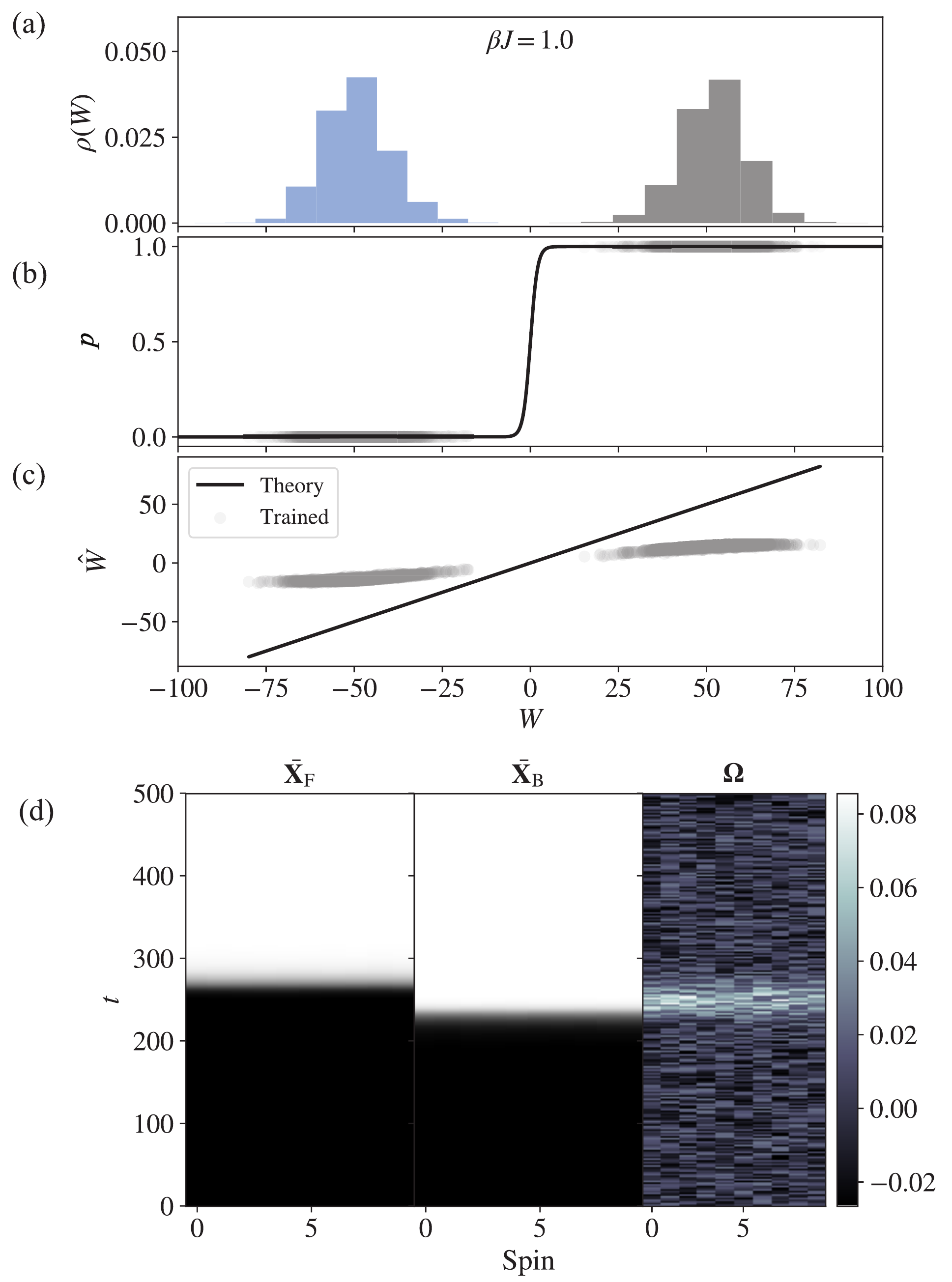}
    \caption{A highly irreversible process. (a) The distributions of forward (grey) and reverse (blue) work are well-separated. (b) The forward likelihood of sample trajectories is either 0 or 1, and the prediction (grey circles, matches the theory (solid curve). (c) In this example, the value of work that the classifier calculates $\hat{W}$ (obtained by inverting the sigmoid function) is different than the actual value of work $W$. (d) The average forward ($\bar{\mathbf{X}}_{\rm{F}}$) and backward ($\bar{\mathbf{X}}_{\rm{B}}$) trajectories, and the network wights $\mathbf{\Omega}$ suggests that the spin orientations midway through the process is a way to decide the direction of time's arrow.     }
    \label{fig:FigA4}
\end{figure}

\end{document}